\documentclass{elsart41}

\usepackage{amsfonts}
\usepackage{amsmath}
\usepackage{amssymb}
\usepackage{graphicx}

\begin{document}

\begin{frontmatter}

\title{Study of the spin-$\frac32$ Hubbard-Kondo lattice model by means of the Composite Operator
Method}

\author{A. Avella\corauthref{Avella}}
\and
\author{F. Mancini}

\address{Dipartimento di Fisica ''E.R. Caianiello" - Laboratorio
Regionale SuperMat, INFM Universit\`{a} degli Studi di Salerno,
I-84081 Baronissi (SA), Italy}

\corauth[Avella]{Corresponding author. Tel: +39 089 965418. Fax: +39
089 965275. E-mail: avella@sa.infn.it. Homepage:
http://www.sa.infn.it/Homepage.asp?avella}

\begin{abstract}
We study the spin-$\frac32$ Hubbard-Kondo lattice model by means of
the Composite Operator Method, after applying a Holstein-Primakov
transformation. The spin and particle dynamics in the ferromagnetic
state are calculated by taking into account strong on-site
correlations between electrons and antiferromagnetic exchange among
$\frac32$ spins, together with usual Hund coupling between electrons
and spins.
\end{abstract}

\end{frontmatter}

The revival in the study of manganites has led to experimental
re-examination of their different properties. One of the puzzling
features is the strong deviation of the spin-wave dispersion from
the typical Heisenberg behavior. In particular, it has been observed
an unexpected softening at the zone boundary \cite{Hwang_98}. These
observations are very important as they indicate that some aspects
of spin dynamics in manganites have not been entirely understood yet
\cite{Khaliullin_00,Mancini_00e}. According to this, we have decided
to investigate the spin dynamics of the ferromagnetic state of the
spin-$\frac32$ Hubbard-Kondo lattice model by means of the Composite
Operator Method (COM) \cite{Mancini_04}. The Hamiltonian under
analysis reads as
\begin{align}
H&=\sum_{\mathbf{ij}}\left(
-2dt\alpha_{\mathbf{ij}}-\mu\delta_{\mathbf{ij}}\right)
c^{\dagger}\left(  i\right) c\left(  j\right) + U\sum
_{\mathbf{i}}n_{\uparrow}\left(  i\right) n_{\downarrow}\left(
i\right) \nonumber \\
& -J_{\mathrm{H}}\sum_{\mathbf{i}}\mathbf{s}\left( i\right) \cdot
\mathbf{S}\left(  i\right) +dJ_{AF}\sum_{\mathbf{i}}\mathbf{S}\left(
i\right) \cdot\mathbf{S}^{\alpha}\left(  i\right)
\end{align}
where $\mathbf{i}$ is a vector of the $d$-dimensional lattice and
$i=({\bf i},t)$, $\mu$ is the chemical potential, $c^{\dag
}(i)=\left( c_{\uparrow }^{\dag }(i),c_{\downarrow }^{\dag }(i)
\right)$ is the electronic creation operator in spinorial notation,
$t$ is the hopping amplitude, $\alpha_{\mathbf{ij}}$ is the
nearest-neighbor projector,  $U$ is the on-site Coulomb interaction,
$n_{\sigma}\left(  i\right) = c_{\sigma}^{\dagger}\left( i\right)
c_{\sigma}\left(  i\right)$, $J_{\mathrm{H}}$ is the Hund coupling,
$\mathbf{s}\left( i\right) =\frac{1}{2}c^{\dagger}\left( i\right)
\boldsymbol{\sigma} c\left( i\right)$, $\boldsymbol{\sigma}$ are the
Pauli matrices, $\mathbf{S}\left( i\right) $ is the core
$\frac32$-spin, $J_{AF}$ is the antiferromagnetic coupling. We have
used the notation $\phi^{\alpha}\left(  i\right)
=\sum_{\mathbf{j}}\alpha_{\mathbf{ij} }\phi\left( \mathbf{i},t
\right)$.

In order to avoid the difficulties related to the high value of the
core spin, we have used the Holstein-Primakoff transformation:
$S_{+}\left(  i\right) = \sqrt{2S-n_{a}\left( i\right)  }a\left(
i\right)$, $S_{-}\left( i\right) = a^{\dagger}\left(  i\right)
\sqrt{2S-n_{a}\left( i\right) }$, $S_{z}\left(  i\right) =
S-n_{a}\left(  i\right)$, where $S=\frac32$, $a\left( i\right)  $ is
a spinless bosonic destruction operator and $n_{a}\left( i\right) =
a^{\dagger }\left(  i\right) a\left( i\right) $. Then, we have
decided to approximate the non-linear term $\sqrt{2S-n_{a}\left(
i\right)  }$ to the first order in $\frac{\delta n_{a}\left(
i\right)  }{2S-n_{a}}$ where $\delta n_{a}\left(  i\right)
=n_{a}\left( i\right)  -\left\langle n_{a}\left(  i\right)
\right\rangle $ and $n_{a}=\left\langle n_{a}\left( i\right)
\right\rangle $. It is worth noticing that this approximation
preserves all properties related to the angular momentum algebra of
the core spin. The transformed Hamiltonian reads as
\begin{align}
H & =\sum_{\mathbf{ij}}\left( -2dt\alpha_{\mathbf{ij}}-\mu\delta
_{\mathbf{ij}}\right)  c^{\dagger}\left(  i\right) c\left( j\right)
+U\sum_{\mathbf{i}}n_{\uparrow}\left(  i\right) n_{\downarrow}\left(
i\right) \nonumber \\ & -J_{\mathrm{H}}\sum_{\mathbf{i}}\left[
S-n_{a}-\delta n_{a}\left( i\right)  \right]  s_{z}\left(  i\right)
\nonumber
\end{align}
\begin{align}
& -\frac{1}{2}J_{\mathrm{H}} A\sum_{\mathbf{i}}\left[  s_{+}\left(
i\right)  a^{\dagger}\left( i\right) \left(  1-\frac{\delta
n_{a}\left(  i\right) }{2A^{2}}\right)  +h.c.\right]
\nonumber\\
& -2dJ_{AF}\left(  S-n_{a}\right)  \sum_{\mathbf{i}}\delta
n_{a}\left( i\right)\nonumber\\
& +dJ_{AF}A^{2}\sum_{\mathbf{i}}a^{\dagger}\left( i\right)
a^{\alpha}\left(  i\right)
\end{align}
where $A=\sqrt{2S-n_{a}}$.

Within the framework of the COM, we have chosen two operatorial
basis to study the spin and particle dynamics
\begin{align}
& B\left(  i\right)  =\left(
\begin{array}
[c]{l}
a\left(  i\right) \\
s_{+}\left(  i\right)
\end{array}
\right) \hspace{1cm} \psi\left(  i\right)  =\left(
\begin{array}
[c]{l}
\xi_{\uparrow}\left(  i\right) \\
\eta_{\uparrow}\left(  i\right) \\
\xi_{\downarrow}\left(  i\right) \\
\eta_{\downarrow}\left(  i\right)
\end{array}
\right)
\end{align}
where $\xi
\left(i\right)=\left[1-n\left(i\right)\right]c\left(i\right)$ and
$\eta\left(i\right)=n\left(i\right)c\left(i\right)$.

Then, we have linearized the equations of motion by projecting the
currents onto the basis
\begin{align}
& \mathrm{i}\frac{\partial }{\partial t}B\left( \mathbf{i},t\right)
\cong \sum_\mathbf{j}
\varepsilon_{B}\left(\mathbf{i},\mathbf{j}\right) B\left(
\mathbf{j},t\right) \\
& \mathrm{i}\frac{\partial }{\partial t}\psi\left(
\mathbf{i},t\right) \cong \sum_\mathbf{j}
\varepsilon_{F}\left(\mathbf{i},\mathbf{j}\right) \psi\left(
\mathbf{j},t\right)
\end{align}
where
\begin{align}
\varepsilon_{B,F}\left(\mathbf{i},\mathbf{j}\right) & =\sum_{\mathbf{l}}
m_{B,F}\left(\mathbf{i},\mathbf{l}\right)
I_{B,F}^{-1}\left(\mathbf{l},\mathbf{j}\right)\\
m_B\left(\mathbf{i},\mathbf{j}\right) &
=\left\langle\left[\mathrm{i}\frac{\partial }{\partial t}B\left(
\mathbf{i},t\right)B^\dagger\left(
\mathbf{j},t\right)\right]\right\rangle\\
m_F\left(\mathbf{i},\mathbf{j}\right) &
=\left\langle\left\{\mathrm{i}\frac{\partial }{\partial t}\psi\left(
\mathbf{i},t\right)\psi^\dagger\left(
\mathbf{j},t\right)\right\}\right\rangle \\
I_B\left(\mathbf{i},\mathbf{j}\right) & =\left\langle\left[B\left(
\mathbf{i},t\right)B^\dagger\left(
\mathbf{j},t\right)\right]\right\rangle \\
I_F\left(\mathbf{i},\mathbf{j}\right) &
=\left\langle\left\{\psi\left( \mathbf{i},t\right)\psi^\dagger\left(
\mathbf{j},t\right)\right\}\right\rangle
\end{align}
This procedure assures that the neglected component of the current
is orthogonal to the chosen basis. According to this, we have
obtained the corresponding retarded Green's functions in the pole
approximation
\begin{equation}
G_{B,F}\left(  \omega,\mathbf{k}\right)=
\sum_{i}\frac{\sigma_{B,F}^{\left(  i\right)  }\left(
\mathbf{k}\right)  }{\omega-E_{B,F}^{\left(  i\right)  }\left(
\mathbf{k} \right)+\mathrm{i}\delta}
\end{equation}
where the energies $E_{B,F}^{\left(  i\right)  }\left( \mathbf{k}
\right)$ are the eigenvalues of the energy matrices
$\varepsilon_{B,F}\left(\mathbf{k}\right)=\mathcal{F}[\varepsilon_{B,F}\left(\mathbf{i},\mathbf{j}\right)]$
and the spectral densities $\sigma_{B,F}^{\left(  i\right)  }\left(
\mathbf{k}\right)$ can be computed in terms of the normalization
matrices $I_{B,F}$ and of the eigenvectors of the energy
matrices\cite{Mancini_04}. $\mathcal{F}$ is the Fourier transform.

The parameters appearing in the expressions of $m_{B,F}$ and
$I_{B,F}$ have the following definitions
\begin{small}
\begin{align}
n & = \left\langle n\left(  i\right)  \right\rangle =2-\left(
C_{F11}+C_{F22}+C_{F33}+C_{F44}\right)\\
m & = \left\langle s_{z}\left(  i\right)  \right\rangle =
\frac{1}{2}\left( C_{F44}-C_{F22}\right)\\
n_{a} & =C_{B11}-1 \\
\tilde{p}_{1} & = \frac {1}{m}\left(
C_{F11}^{\alpha}+C_{F22}^{\alpha}+C_{F33}^{\alpha}+C_{F44}^{\alpha
}\right)\\
\tilde{p}_{2} &= \frac{1}{A}\frac{\left\langle a\left(  i\right)
s_{-}\left(  i\right)  \right\rangle
}{m}=\frac{1}{A}\frac{C_{B12}}{m}\\
\tilde{p}_{3} & =\frac{1}{A^{2}}\frac{\left\langle \delta
n_{a}\left( i\right)
s_{z}\left(  i\right)  \right\rangle }{m}\\
\tilde{p}_{4} & =\frac{1}{A^3}\frac{\left\langle \delta n_{a}\left(
i\right)  a\left( i\right)
s_{-}\left(  i\right)  \right\rangle}{m}\\
\tilde{p}_{5} & = \frac {1}{A^{2}}\frac{\left\langle \delta
n_{a}\left(  i\right)  n\left( i\right) \right\rangle }{m} \\
\Delta_{0} & =\frac{1}{2}\left(  C_{F11}^{\alpha}-C_{F22}^{\alpha}
+C_{F33}^{\alpha}-C_{F44}^{\alpha}\right) \\
\Delta_{z}& = \frac{1}{2}\left(  C_{F11}^{\alpha}-C_{F22}^{\alpha}
-C_{F33}^{\alpha}+C_{F44}^{\alpha}\right) \\
p & =\left\langle \left(\frac{1}{4}n^{\alpha}\left(  i\right)
n\left( i\right) + \mathbf{s}^{\alpha}\left( i\right) \cdot
\mathbf{s}\left( i\right) \right) \right\rangle  \nonumber\\
&-\left\langle \left( c_{\uparrow}\left( i\right)
c_{\downarrow}\left(  i\right) \right) ^{\alpha}c_{\downarrow
}^{\dagger}\left(  i\right)
c_{\uparrow}^{\dagger}\left(  i\right) \right\rangle \\
\chi_{z}^{\alpha} & =\left\langle n^{\alpha}\left(  i\right)
s_{z}\left( i\right)  \right\rangle
\end{align}
\end{small}
where $C_{F\beta\gamma} =\left\langle \psi_\beta\left(
\mathbf{i}\right) \psi_\gamma^\dagger\left(  \mathbf{i}\right)
\right\rangle $, $C_{F\beta\gamma}^\alpha =\left\langle
\psi^\alpha_\beta\left( \mathbf{i}\right) \psi_\gamma^\dagger\left(
\mathbf{i}\right) \right\rangle $ and $C_{B\beta\gamma}
=\left\langle B_{\beta}\left( \mathbf{i}\right)
B_{\gamma}^{\dagger}\left( \mathbf{i}\right) \right\rangle$.
According to the prescriptions of the COM \cite{Mancini_04}, the
parameters that cannot be computed by their definitions
($\tilde{p}_{3}$, $\tilde{p}_{4}$, $\tilde{p}_{5}$, $p$ and
$\chi_{z}^{\alpha}$) would be fixed by the following relations
\begin{align}
C_{11} & =C_{33} \quad \quad C_{12}  =0 \\
C_{34} & =0 \quad \quad C_{44}  =C_{B22}
\end{align}
coming from the algebra and by one more relation coming from the
request that the hydrodynamic limit should be satisfied (i.e., the
existence of a sound mode). All these equations (definitions and
constraints) form a coupled system that should be computed
self-consistently. The results of these calculations will be
presented elsewhere.

In conclusion, we have reported the solution for the Hubbard-Kondo
model in presence of antiferromagnetic coupling between the core
spin with the framework of the Composite Operator Method in the pole
approximation. The model has been first mapped through the
Holstein-Primakov transformation that has been then approximated to
the first order in the number fluctuation operator. The spin
dynamics has been fully determined and will be analyzed numerically.


\begin{thebibliography}{1}
\bibitem{Hwang_98} H.~Y. Hwang, et~al., Phys.~Rev.~Lett. 80 (1998) 1316.
\bibitem{Khaliullin_00} G.~Khaliullin, P.~Kilian, Phys.~Rev.~B 61 (2000) 3494.
\bibitem{Mancini_00e} F.~Mancini, N.~B. Perkins, N.~M. Plakida, Phys.~Lett.~A 284 (2001) 286.
\bibitem{Mancini_04} F.~Mancini, A.~Avella, Adv.~Phys. 53 (2004) 537.
\end{thebibliography}

\end{document}